# Satisfying the Direct Laser Acceleration Resonance Condition in a Laser Wakefield Accelerator

J. L. Shaw[1,a)], N. Vafaei-Najafabadi[1], K. A. Marsh[1], N. Lemos[1], F. S. Tsung[1], W. B Mori[1], and C. Joshi[1]

[1]*University of California Los Angeles, Los Angeles CA USA 90095*

a)*Corresponding author: jshaw05@ucla.edu*

**Abstract.** In this proceeding, we show that when the drive laser pulse overlaps the trapped electrons in a laser wakefield accelerator (LWFA), those electrons can gain energy from direct laser acceleration (DLA) over extended distances despite the evolution of both the laser and the wake. Through simulations, the evolution of the properties of both the laser and the electron beam is quantified, and then the resonance condition for DLA is examined in the context of this change. We find that although the electrons produced from the LWFA cannot continuously satisfy the DLA resonance condition, they nevertheless can gain a significant amount of energy from DLA.

## INTRODUCTION

In a laser wakefield accelerator (LWFA) [1], direct laser acceleration (DLA) [2] [3] can contribute significantly to the final observed energy of the electrons as long as the drive laser overlaps the electrons that are trapped in the first bucket of the wake [4]. DLA can occur in such a system because the betatron oscillation of the electrons in the plane of the laser polarization (in the presence of an ion column) can lead to energy transfer from the transverse laser field to the transverse motion of the electrons. This increased transverse momentum can then be converted to increased longitudinal momentum by coupling to the magnetic field of the laser. The resonance condition for DLA is given by [2] [3]:

$$N\omega_\beta = (1 - v_\parallel/v_\phi)\omega_0 \qquad (1)$$

where $\omega_\beta = \omega_p/(2\gamma)^{1/2}$ is the betatron frequency of the electron, N is an integer indicating the harmonic of the betatron frequency, $v_\parallel$ is the longitudinal velocity of the electron, $v_\phi$ is the phase velocity of the laser, and $\omega_0$ is the laser frequency. Essentially, this resonance condition means that in order for an electron to gain energy from DLA, a harmonic of the betatron frequency $N\omega_\beta$ must equal the Doppler-shifted laser frequency $(1-v_\parallel/v_\phi)\omega_0$ witnessed by the electron [2].

It is well-known that in LWFAs, especially those not in the ideal blowout regime [5], the properties of the drive laser, including $\omega_0$ and $v_\phi$ [6], evolve. Furthermore, as electrons are accelerated in a LWFA, their longitudinal momentum, and therefore their $v_\parallel$, increases, and their betatron frequency is expected to fall as $\gamma^{-1/2}$. Given that the four quantities present in Equation 1 are all evolving, it would appear unlikely that the DLA resonance condition could be satisfied long enough for electrons in a LWFA to gain net energy from DLA. In this proceeding, we demonstrate through particle-in-cell (PIC) simulations of LWFA-produced electrons that it is not necessary to continuously satisfy Equation 1 for these electrons to gain sizable energy from DLA over extended distances.

## OSIRIS SIMULATION OF A LWFA WITH DLA

To investigate how electrons in a LWFA can gain energy from DLA over extended distances, we examined a two-dimensional OSIRIS [7] PIC-code simulation of a LWFA where the produced electrons show energy gain from DLA [4]. For the simulation, the laser pulse has a normalized vector potential $a_0 = eA/mc^2$ of 2.1 and a central wavelength of 815 nm. Its pulse length was 45 fs so that it overlapped the electrons trapped in the first bucket of the wake. The laser ionized a neutral gas, which was comprised of 99.9% helium and 0.1% nitrogen so that the inner-shell nitrogen electrons would be trapped via the ionization injection mechanism [8] [9]. The resulting plasma density profile consisted of an 1800-μm-long plateau region with 100-μm-long linear entrance and exit ramps. The simulation was done in a speed-of-light moving window and used a 2494 x 452 grid. The longitudinal and transverse resolutions were $k_0 dx_\parallel = 0.21$ and $k_p dx_\perp = 0.12$, respectively. The simulation was run to completion once to identify the highest-energy electrons it produced. Those electrons were tagged, and the simulation was then rerun to determine the position and momenta of those electrons at each time step in the simulation. With this information, it was possible to calculate the relative contribution of the wakefield to the total energy of the electrons using

$$\int_0^t \vec{E}_\parallel \cdot \vec{v}_\parallel dt, \tag{2}$$

where $\vec{E}_\parallel$ is the longitudinal electric field sampled by the electrons. Similarly, the relative contribution due to DLA was calculated using

$$\int_0^t \vec{E}_\perp \cdot \vec{v}_\perp dt, \tag{3}$$

where $\vec{E}_\perp$ is the transverse electric field at the location of these electrons, which is dominated by the electric field of the laser, and $\vec{v}_\perp$ is their transverse velocity. The highest-energy electrons reached a max energy of 290 MeV by the end of the simulation. Using Equations 2 and 3, it was determined that 197 MeV of the 290 MeV was due to LWFA and 93 MeV was due to DLA. Net energy gain from DLA occurred continuously for 1180 μm and included nearly 4 complete betatron oscillations of the electrons. In other words, for this LWFA configuration, the contribution of DLA to the final electron energy is comparable to the wakefield contribution [4]. One of these electrons, which gained 38% of its total energy from DLA, was randomly selected, and this test electron will be used in all subsequent calculations to investigate the DLA resonance condition in such a LWFA.

## MEASURING EVOLVING QUANTITIES IN A LWFA

In order to determine if this test electron satisfies the DLA resonance condition for the regions where it gained energy from DLA, the evolution of its relevant parameters was followed throughout the simulation. The betatron frequency was measured by first plotting the test electron's trajectory, which shows its betatron oscillations, as a function of distance in the simulation. From this plot, the locations in the simulation where the electron reaches a minimum, reaches a maximum, or crosses a zero (i.e. the laser axis) of its betatron oscillation can be determined. Once the locations are known for each of the minima, maxima, or zeros, the instantaneous half betatron wavelength $\lambda_\beta/2$ at each of these points can be measured by taking the distance between the points $\lambda_\beta/4$ before and after a given point. For example, the test electron reached a zero of its betatron oscillation at 396 μm; $\lambda_\beta/2$ is estimated as 75 μm at this position by taking the difference between the locations of the next minimum in the betatron oscillation (438 μm) and of the prior maximum (363 μm). $\lambda_\beta$ was then converted to $\omega_\beta$. This measured $\omega_\beta$ does not fall as predicted by the betatron frequency $\omega_{\beta,expected} = \omega_p/(2\gamma)^{1/2}$ expected for an electron in an ion column as seen in Fig. 1.

Since $\gamma$ and the longitudinal momentum $\vec{p}_\parallel$ are known as a function of distance from tracking the test electron through the simulation, $\vec{v}_\parallel$ can be directly calculated using $\vec{p}_\parallel = \gamma m_e \vec{v}_\parallel$. As Fig. 1 shows, $v_\parallel/c$ (red curve) quickly asymptotes to 1 because the test electron rapidly gains energy after it is born. The phase velocity $v_\phi$ was calculated by measuring the advance of a constant-phase front of the laser cycle at the location of the test electron in each simulation frame [4]. Since the simulation window is moving at the speed of light, if the constant-phase front advances with respect to the simulation window, it indicates a phase velocity that is greater than the speed of light. This $v_\phi$ is expected from the plasma dispersion relation since there is residual plasma electron density on axis. Figure 1 shows that $v_\phi/c -1$ (black curve) decreases with propagation distance because the percentage of blowout is increasing. The quantity $v_\phi/c -1$ is plotted instead of $v_\phi/c$ to emphasize the evolution of the phase velocity since Equation 1 is very sensitive to the quantity. The discrete nature of this curve is due to the method of measurement. Since the measurement considered the phase velocity of the laser cycle at the location of the electron, the average $\Delta z$ of the phase front relative to the moving window can be measured for the $\Delta t$ over which the test electron samples that particular cycle. That gives an average $v_\phi$ for one sampling period. However, when the electron moves to the next laser cycle, the continuity of the measurement is broken. The $v_\phi$ measurement must start again, which yields discrete jumps in the sampled $v_\phi$. Finally, the laser frequency was calculated by taking the on-axis lineout of the laser at each simulation frame, measuring the wavelength of the laser cycle at the location of the electron in that frame, and converting that wavelength to a frequency using $\omega_0 = 2\pi v_\phi/\lambda$. The discrete nature of the measurement of the laser wavelength is caused by the resolution (i.e. simulation grids per laser wavelength) of the simulation. Overall, it is clear from Fig. 1 that all four quantities ($\omega_\beta$, $v_\parallel$, $\omega_0$, and $v_\phi$) present in the DLA resonance condition evolve significantly in a LWFA. The question that remains is whether that resonance condition holds while these quantities evolve.

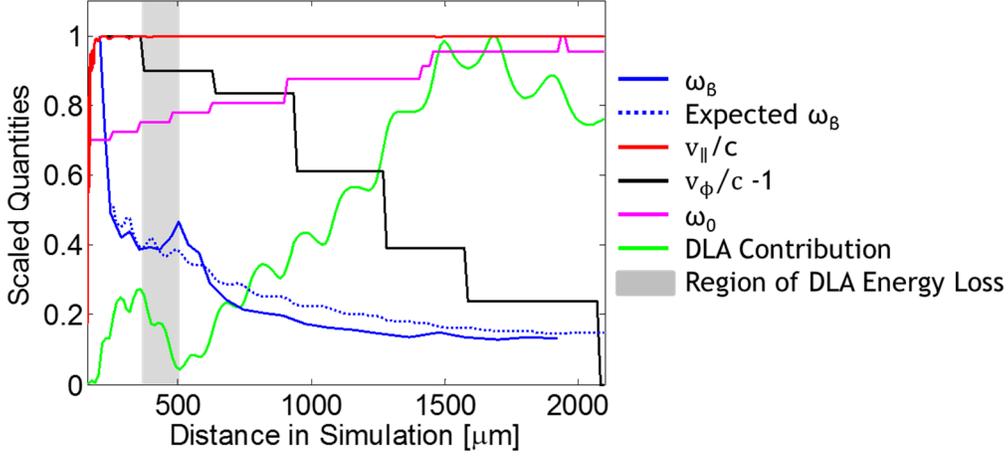

**FIGURE 1.** Plot of the scaled quantities $\omega_\beta$ (solid blue curve), $v_\parallel/c$ (red curve), $v_\phi/c - 1$ (black curve), and $\omega_0$ (magenta curve) as a function of propagation distance in the simulation. $\omega_\beta$, $\omega_0$, and $v_\phi/c -1$ are scaled to their maximum values $3.1 \times 10^{13}$ Hz, $3.3 \times 10^{15}$ Hz, and 0.0017, respectively. The plot starts at 160 μm into the simulations, which is where the test electron was born and which is located at the start of the constant-density region of the plasma. Also plotted for reference is the DLA energy contribution $\int_0^t \vec{E}_\perp \cdot \vec{v}_\perp dt$ (green curve) scaled to its maximum value of 93 MeV. For comparison, the expected $\omega_\beta$ (blue dotted curve) is also plotted after being scaled to $3.1 \times 10^{13}$ Hz, which was the same value used to normalize the measured $\omega_\beta$. The region over which the test electron first loses net energy to the transverse laser field is shaded gray.

## TESTING THE DLA RESONANCE CONDITION WITH EVOLVING QUANTITIES

Though all the quantities present in the DLA resonance condition are individually evolving, DLA energy gain depends on how they change relative to each other. The evolution of $v_\parallel$, $\omega_0$, and $v_\phi$ all affect the down-shifted laser frequency witnessed by the electron, which is given by the right-hand side of Equation 1 as $(1 - v_\parallel/v_\phi)\omega_0$. This quantity, which was calculated using the measured values shown in Fig. 1, is plotted (red curve) in Fig. 2. This frequency oscillates over the duration of the simulation and is suppressed where the test electron first loses energy to the transverse laser field (shaded region). The value of the down-shifted laser frequency using the measured parameters from Fig. 1 is validated by a direct measurement of the down-shifted laser frequency $\omega_{0,frame}$ in the frame of the test electron (black curve in Fig. 2). $\omega_{0,frame}$ was directly measured by finding the phase of the test electron relative to the peak of the laser cycle at the electron's location in at each frame of the simulation and then using $\omega_{0,frame} = d\phi/dt$.

Seeing that the right-hand side of Equation 1, $(1 - v_\parallel/v_\phi)\omega_0$, accurately predicts the down-shifted laser frequency witnessed by the electron, it can then be compared to the left-hand side of Equation 1, $N\omega_\beta$, to see if the DLA resonance condition is satisfied where the test electron is gaining net energy from DLA. For this comparison, the value N=1 is taken because the electron slips roughly one laser cycle per betatron oscillation in this simulation.

Comparing the values of $N\omega_\beta$ and $(1-v_\parallel/v_\phi)\omega_0$ in Fig. 2 shows that right after the electron is born, both the betatron frequency and the down-shifted laser frequency sampled by the test electron are rapidly changing, which makes a one-to-one comparison difficult in this region where the test electron rapidly gains energy from DLA. However, shortly after this initial energy gain, the test electron experiences a rapid energy loss to the transverse laser field. In this region, not only does the betatron frequency stop falling, but also the down-shifted laser frequency is reduced significantly; as expected, the DLA resonance condition cannot be satisfied in this region. However, after this first and rapid loss of energy to the transverse laser field, the test electron again begins gaining energy from DLA for an extended duration. During this net energy gain from DLA, one would expect the left-hand side (blue curve) and right-hand side (red curve) of the DLA resonance condition to agree. However, this is not the case. Instead, the down-shifted laser frequency only agrees intermittently with the betatron frequency. Even though the test electron does not exactly satisfy the DLA resonance condition, it gains substantial energy over nearly four complete betatron oscillations. How this electron gains energy from DLA while only intermittently satisfying the resonance equation must now be explained.

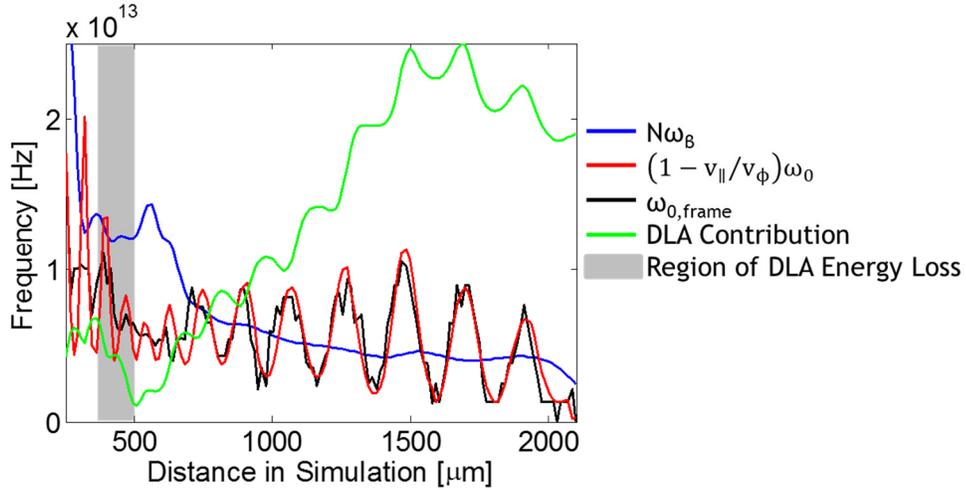

**FIGURE 2.** $N\omega_\beta$ (blue curve) and $(1-v_\parallel/v_\phi)\omega_0$ (red curve) calculated for the test electron as a function of the distance into the simulation. Both curves were calculated using the values measured in Fig. 1. Also shown (black curve) is the laser frequency $\omega_{0,frame}$ witnessed by the electron in its frame. The green curve is the DLA contribution to the total energy gain scaled to its maximum value, and the shaded region marks where the electron first loses net energy to the transverse laser field.

## DLA ENERGY GAIN WITH PERIODIC RESONANCE

The test electron can be used to investigate how DLA energy gain occurs even if the test electron can only intermittently satisfy the resonance condition. Energy gain occurs when the transverse laser field and the transverse momentum of the electron have opposite signs. Therefore, since both $p_\perp$ and the transverse field are known at each frame in the simulation, the locations where the test electron gains energy from the transverse laser field can be determined. Even though the resonance condition is not satisfied continuously where the electron gains net energy from DLA, the test electrons can gain this energy over an extended distance because the transverse momentum (blue curve in Fig. 3) of the test electron and the transverse laser field (red curve in Fig. 3) evolve such that the electron is in an accelerating phase (blue dots in Fig. 3) for more than half of each betatron oscillation.

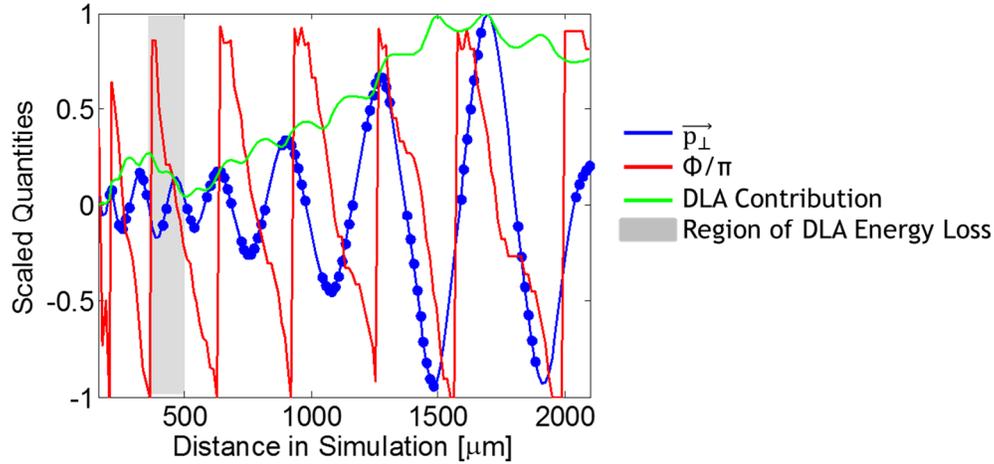

**FIGURE 3.** Plot of the phase ϕ/π (red curve) of the test electron relative to the sampled laser cycle and of the transverse momentum (blue curve) of the test electron scaled to its maximum value $p_\perp/mc = 36.4$. The blue dots along the transverse momentum curve indicate the simulation frames where the test electron is gaining energy from DLA. The discrete jumps in the phase curve mark where the test electron slipped back one laser cycle. The green curve is the DLA contribution to the energy gained by the test electron scaled to its maximum value, and the shaded area indicates where the test electron first experiences net energy loss to the transverse laser field.

## CONCLUSION

In a LWFA system where both the properties of the electrons and the properties of the laser are evolving, trapped electrons can gain energy from DLA as long as the drive laser overlaps the trapped electrons. This proceeding has shown that though such electrons at best can only satisfy the DLA resonance condition intermittently, they gain net energy from DLA in such regions because they are accelerated by the transverse laser field for more than one-half of each betatron cycle. This understanding of DLA in LWFA systems will allow DLA to be exploited in future LWFA experiments where the maximum energy of the produced electron beams is the primary outcome.

## ACKNOWLEDGMENTS

Experimental work supported by DOE grant DE-SC0010064. Simulation work done on the Hoffman2 Cluster at UCLA. Fellowship provided by NSF Graduate Fellowship DGE-0707424.

## REFERENCES

1. T. Tajima and J. M. Dawson, Phys. Rev Lett. **43**, 267 (1979).
2. A. Pukhov, Reports on Progress in Physics **66**, 47 (2003).
3. A. Pukhov, Z. –M. Sheng, and J. Meyer-ter-Vehn, Phys. Plasmas **6**, 2847 (1999).
4. J. L. Shaw, F. S. Tsung, N. Vafaei-Najafabadi, K. A. Marsh, N. Lemos, W. B. Mori, and C. Joshi, Plasma Phys. Contr. F. **56**, 084006 (2014).
5. W. Lu, M. Tzoufras, C. Joshi, F. S. Tsung, W. B. Mori, J. Vieira, R. A. Fonseca, and L. O. Silva, Phys. Rev. Spec. Top. – Accel. Beams **10**, 061301 (2007).
6. W. B. Mori, IEEE J. Quantum Electron. **33**, 1942 (1997).
7. R. A. Fonseca et al., in *Lecture Notes in Computer Science*, Vol. 2331 (Springer Berlin Heidelberg, 2002) pp. 342-351.
8. A. Pak, K. A. Marsh, S. F. Martins, W. Lu, W. B. Mori, and C. Joshi, Phys. Rev. Lett. **104**, 025003 (2010).
9. C. McGuffey et al., Phys. Rev. Lett. **104**, 025004 (2010).